# High-Q Multimodal Guided Surface Lattice Resonances in Index-Discontinuous Environments

Suichu Huang[1, 2], Kan Yao[3], Hao Wang[1], Xumin Ding[2,*], Feiran Li[1], Cong Huang[1], Wentao Huang[1], Xuezheng Zhao[1, *], Yuebing Zheng[3, *], Yunlu Pan[1,*]

1 State Key Laboratory of Robotics and Systems, School of Mechatronics Engineering, Harbin Institute of Technology, Harbin 150001, China

2 Advanced Microscopy and Instrumentation Research Center, School of Instrumentation Science and Engineering, Harbin Institute of Technology, Harbin, 150080, China

3 Walker Department of Mechanical Engineering, Material Science and Engineering Program and Texas Materials Institute, The University of Texas at Austin, Austin TX 78712, United States

* Corresponding author. Email: xuminding@hit.edu.cn, zhaoxz@hit.edu.cn, zheng@austin.utexas.edu, yunlupan@hit.edu.cn

**Abstract**

Surface lattice resonances (SLRs) in metasurfaces have become a transformative platform for subwavelength optical devices, leveraging their high quality (Q)-factors, pronounced local field enhancement, and extensive long-range interactions. However, current high-Q SLR implementations are fundamentally limited by their dependence on homogeneous dielectric environments. This restriction significantly hinders their applicability in emerging fields such as molecular sensing, where operation in heterogeneous dielectric media (e.g., interfaces with an aqueous or air cladding) is often indispensable. To overcome this limitation, we introduce guided surface lattice resonances (gSLRs) by integrating nanoparticle arrays within slab waveguides. This configuration facilitates efficient coupling between scattered light and Bloch modes, thereby enabling high-Q multimodal resonances even in index-discontinuous environments. Experimental validation under incoherent illumination demonstrates a Q-factor of 1489 in an index-mismatched surrounding. Furthermore, the coupling strength and resonance intensity of these multimodal gSLRs can be continuously modulated by adjusting the vertical displacement of the nanoparticle arrays within the slab layers. To augment the sensitivity to local dielectric variations, we investigate gSLRs in metasurfaces integrated with metallic substrates, demonstrating suitability for biomolecule detection. A mathematical sensing model, incorporating biochemical reaction kinetics and optical responses, is established by representing adsorbed molecules as a uniform dielectric layer and validated through bovine serum albumin (BSA) sensing experiments. This work not only advances the fundamental understanding of resonance engineering in complex media but also facilitates the development of ultrathin, ultra-compact nano-optical and optoelectronic devices.

## Introduction

Plasmonic nanostructures have garnered considerable interest owing to their capacity to effectively confine light at sub-wavelength scales and substantially enhance local fields[1], thereby enabling a range of applications including energy conversion[2,3], sensing[4,5], photodetecting[6] and spectroscopy[7-9]. A significant impediment to the development of high-performance devices is the low quality (Q)-factor of resonances in individual plasmonic structures, which is attributed to the intrinsic loss of plasmonic materials at optical frequencies[10]. To tackle this issue, two-dimensional (2D) nanostructure arrays (metasurfaces) were investigated, wherein collective resonant modes, also referred to as surface lattice resonances (SLRs), were generated, exhibiting exceptionally high Q-factors (high-Q), remarkably narrow linewidths, and significantly enhanced near-fields[11-14]. These exceptional attributes facilitated strong light-matter interactions and reduced the threshold power density, making them highly advantageous for applications in lasing[15-18], strong coupling[19-21] and nonlinear optics[17,22,23]. However, a major drawback of SLRs is the need to excite high-Q resonances in homogeneous or index-continuous environments[12,24,25]. This requirement is challenging to meet in contexts where the structures need to closely interact with external environments, such as in biosensing applications[26-28].

Several attempts have been undertaken to achieve SLRs in index-discontinuous environments, including engineering the geometries of nanoparticles (meta-atoms)[29-31], positioning plasmonic arrays at specific distances inside the substrate[32], and fabricating nanostructure arrays on metal films[33-35]. Although some studies have reported considerably high Q-factors, the underlying mechanisms remain elusive[34,35]. Consequently, designing metasurfaces with high-Q SLRs systematically remains difficult, and numerous SLR-based applications are hindered by the absence of explicit theoretical models. For instance, recent research on biosensors derived from SLRs yielded inconsistent response characteristics[36,37], which were obtained by fitting the results with empirically selected functions instead of rigorous mathematical models. This inconsistency caused ambiguity in explanation of the sensing mechanism, presenting challenges in calibration and optimization. Establishing an approach to rationally designing metasurfaces with high-Q resonant modes in index-inhomogeneous backgrounds is therefor crucial.

In this study, we propose and experimentally validate high-Q gSLRs within index-discontinuous environments, which are realized by embedding plasmonic nanoparticle arrays in slab waveguides and engirneering the radiation patterns of the meta-atoms. A resonance peak with a linewidth as narrow as approximately 0.60 nm and a Q-factor reaching 1489 is achieved. This Q-factor represents the highest reported value for a metasurface with lossy materials under incoherent illumination in the near-visible range[16,38-40]. In addition, the presence of diverse transverse electric (TE) and transverse magnetic (TM) modes imparts multimodal characteristics to gSLRs. Furthermore, the coupling strength and peak intensity of the specific gSLR modes can be regulated by altering the vertical position of the array within the slab. Moreover, the gSLR mode positions are highly sensitive to the dielectric layer thickness, which is suitable for biosensing applications. As a proof-of-concept, we engineer gSLRs in metasurfaces with metallic substrates, achieving enhanced sensitivity for meticulous analysis of the adsorption of biomolecules on metallic surfaces. A mathematical model is constructed by integrating biochemical reaction kinetics with the gSLR theory, which accurately elucidates the experimental results.

This work, by generating high-Q multimodal resonances in index-discontinuous environments and enabling adaptable mode intensity engineering, demonstrates significant potential for applications in white-light lasers, structural color and multiband filters.

## Results

### Design of gSLR metasurfaces

The configuration of proposed metasurface is illustrated in Fig. 1a. It comprises a gold nanosphere array (AuNSA) integrated within the core layer of a three-layered slab waveguide. This waveguide consists of a semi-infinite air superstrate ($n_1$ = 1), a dielectric core layer (SU8 photoresist, $n_2$ = 1.56) with a thickness denoted as $t_{slab}$, and a semi-infinite $SiO_2$ substrate ($n_3$ = 1.45). AuNSA is composed of gold nanospheres (AuNSs) with a diameter of $d$, and the lattice constants are $p_x$ and $p_y$ in the $x$- and $y$-directions, respectively. When configuring a metasurface, the AuNSA can be positioned at any vertical location within the core layer. Initially, we consider a typical scenario in which the AuNSA is located at the bottom of the thin slab, in direct contact with the substrate, as this configuration is relatively straightforward to fabricate. When the core slab is sufficiently thick, the three-layered structure supports guided modes. The in-plane propagation constant of the guided modes $k_{GM}$ can be determined by solving the dispersion equation (Supplementary Note 1).

$$\tan(\beta_2 t) = \frac{\beta_2(c_{21}\beta_1 + c_{23}\beta_3)}{\beta_2^2 - c_{21}\beta_1 c_{23}\beta_3} \tag{1}$$

Here, $\beta_1 = \sqrt{k_{GM}^2 - n_1^2 k_0^2}$, $\beta_2 = \sqrt{n_2^2 k_0^2 - k_{GM}^2}$, $\beta_3 = \sqrt{k_{GM}^2 - n_3^2 k_0^2}$ are out-of-plane wave vectors in the superstrate, core layer, and substrate, respectively, and $n_1$, $n_2$, and $n_3$ are the refractive indices of materials. $c_{21}$ and $c_{23}$ are constants. For TE modes (only the $E_y$-component), $c_{21} = c_{23} = 1$, and for TM modes (only the $H_y$-component), $c_{21} = \frac{n_2^2}{n_1^2}$ and $c_{23} = \frac{n_2^2}{n_3^2}$.

When a single AuNS within the slab waveguide is excited by a light beam, a portion of the scattered light is coupled to the guided modes, propagating with an in-plane wave vector denoted as $k_{GM}$, as depicted in Fig. 1b. Considering an SU8 slab with a thickness of 1200 nm as an example, the slab waveguide supports TE0, TM0, TE1, and TM1 modes within the 750–950 nm wavelength range. Figure S1 delineates the distribution of scattered energy between that which propagates freely into space and that which is confined within the waveguide structure, for a 100 nm diameter AuNS positioned at the base of the slab waveguide. This figure . A significant amount of light is confined within the waveguide. The far-field radiation patterns of guided light under *x*-polarized excitation are shown in Fig. 1g, and they are similar to the in-plane radiation patterns of the electric dipoles (for TE0 and TE1) and magnetic dipoles (for TM0 and TM1). The scattered light that couples to the TE0 and TE1 modes propagates mainly in the *y*-direction, whereas it couples to the TM0 and TM1 modes in the *x*-direction.

When an AuNSA is integrated into a slab waveguide, collective resonances analogous to those observed in classic surface lattice resonances (SLRs) can emerge. Specifically, if the wave vectors of light scattered by the AuNSs and coupled to the guided modes of the waveguide align with the wave vectors of the Bloch modes within the array (Fig. 1c), the scattered light from neighboring AuNSs will arrive at a given AuNS in phase. This constructive interference leads to coherent re-excitation, ultimately driving the formation of collective resonant modes throughout the array.[12,13]. As the present collective resonances originate from couplings involving guided modes, we termed them guided SLRs (gSLRs). For an AuNSA with lattice constants of $p_x = 600$ nm and $p_y = 550$ nm, the peak center wavelengths of the gSLRs can be determined according to wave vector analysis (Fig. 1c). Under normal excitation, the peak center wavelengths of of the first-order of TE modes are near $\lambda = p_y n_{eff}$, whereas those of TM modes are near $\lambda = p_x n_{eff}$, where $n_{eff}$ is the effective index of the corresponding guided modes. Here, a rectangular array configuration is employed to distinctly separate the TE and TM gSLR modes (refer to Fig. S2 for the square-lattice case). Four gSLR modes are predicted within the 750-950 nm wavelength range, which are labeled TE1(0, ±1), TE0(0, ±1), TM1(±1, 0) and TM0(±1, 0), from shorter to longer wavelengths (Fig. 1d). The simulated transmittance spectrum is shown in Fig. 1e. The peaks in the spectrum redshift from theoretical predictions, with TE1(0, ±1) and TE0(0, ±1) modes shifting larger distances compared to the TM1(±1, 0) and TM0(±1, 0) modes. Meanwhile, the TE1(0, ±1) and TE0(0, ±1) modes have wider expansions than the TM1(±1, 0) and TM0(±1, 0) modes. This is due to the fact that more scattered light couples to the TE1 and TE0 modes (Fig. 1g), causing higher coupling strengths for TE1(0, ±1) and TE0(0, ±1) modes[11,41]. Decreasing the AuNS size reduces the shifting distance and peak linewidth as the energies involved in the couplings decline (Fig. S3). The characteristic field distributions of the four gSLR modes are shown in Fig. 1h. The $E_x$ components of the TE1(0, ±1) and TE0(0, ±1) modes exhibit a single period in the *y*-direction. Similarly, the $H_y$ components of the TM1(±1, 0) and TM0(±1, 0) modes also display one period in the *x*-direction. This observation confirms the classification of guided modes and the associated coupling directions.

As the formation of gSLRs involves guided modes, the structural parameters of the slab waveguide have important influences on the gSLRs. Figure 1f shows the transmittance spectra mapping when the core layer thickness increases from 0 to 1200 nm. The theoretical gSLR peak positions are also drawn for comparison. Light strips are caused by Fabry-Perot interference[42]. When the slab thickness is small, there is no sharp peak in the spectra, since the guided modes are cut off (Supplementary Note 1). As the slab thickness increases, gSLR peaks emerge with deep peaks. The TM1(±1, 0) and TM0(±1, 0) mode peaks in the simulation appear at the same thicknesses as the theoretical prediction, while peaks appear at smaller thicknesses than the theoretical results in the TE1(0, ±1) and TE0(0, ±1) mode branches. These peaks are produced by near-field couplings that result from reflections at the slab boundaries[32].

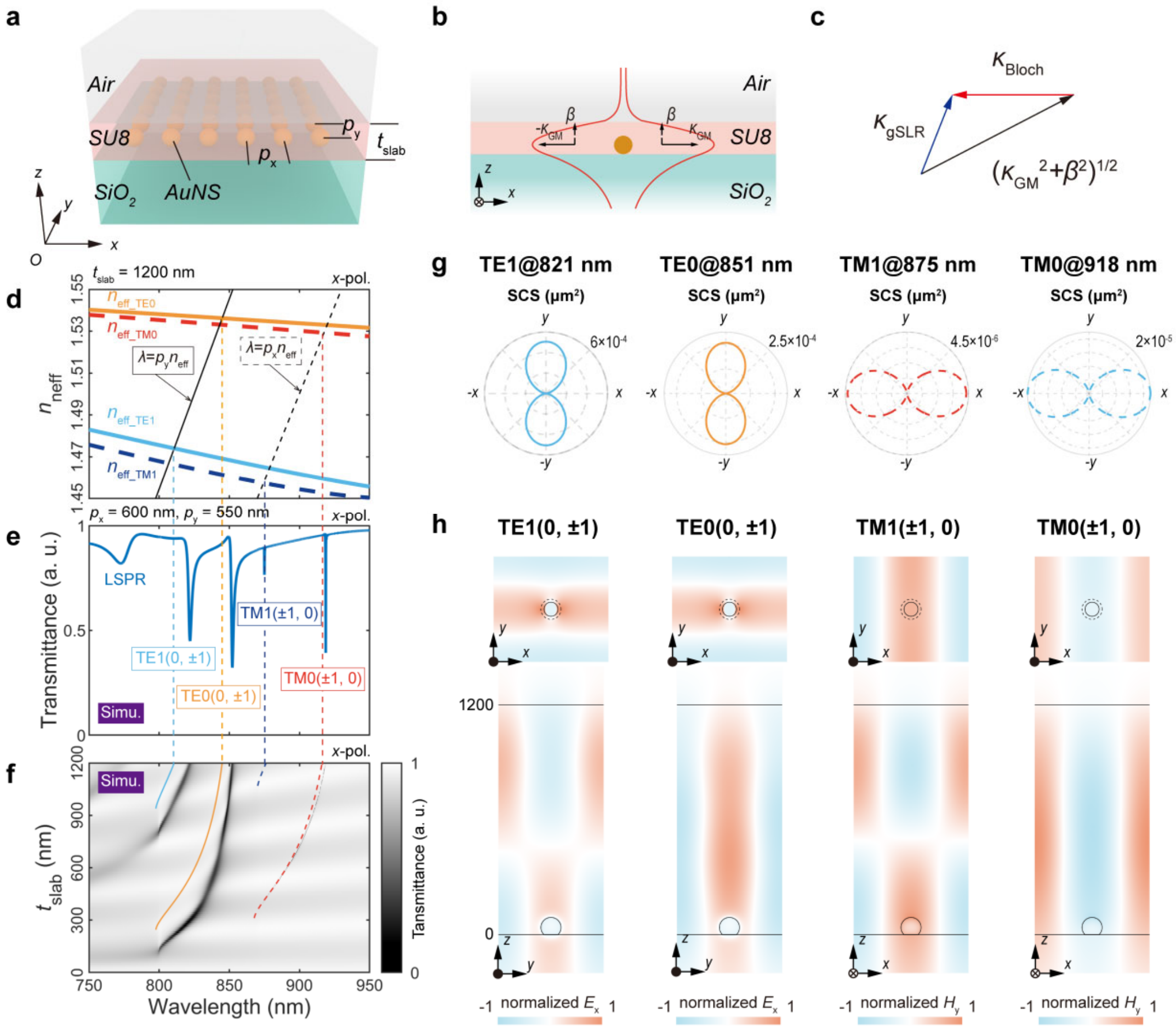

**Figure 1 Principle of gSLRs in index-discontinuous environments. a** Schematic showing a lattice in an index-discontinuous environment. From top to bottom, the index-discontinuous environment involves an air superstrate ($n_1$ = 1), a thin polymer slab (cured SU8 photoresist, $n_2$ = 1.56) with a thickness of $t_{slab}$ and a silica substrate ($n_3$ = 1.45), which compose a slab waveguide. The lattice buried in the bottom of the SU8 slab consists of 100 nm AuNSs with lattice constants of $p_x$ = 600 nm and $p_y$ = 550 nm, respectively. **b** Illustration of scattered light from an AuNS couples to guided modes. $\beta$ and $k_{GM}$ are the out-of-plane and in-plane components of the guided wave vectors. **c** Wave vector analysis of gSLRs. gSLRs are created when scattered light couples to the Bloch momentum $k_{Bloch} = 2\pi/\sqrt{(ip_x)^2+(jp_y)^2}$ introduced by the lattice to generate gSLR in free space, where integers i and j represent coupling order in *x*- and *y*-direction. **d** Determination of gSLR mode peak positions for the slab thickness of 1200 nm. In the range from 750 nm to 950 nm, the slab waveguide supports four guided modes, whose effective indices are plotted by cures, from top to bottom being TE0 (orange solid), TM0 (red dashed), TE1 (light blue solid) and TM1 (dark blue dashed), respectively. The black solid/dashed line indicates the first order coupling in *x*- and *y*-direction. The intersections of solid/dashed lines and cures determine theoretical positions of gSLR modes. Vertical dashed lines guide these mode positions to **e** and **f**. **e** Simulated transmittance spectrum of a lattice in a 1200 nm thick slab.

The four gSLR modes are found to be TE1(0, ±1), TE0(0, ±1), TM1(±1, 0), and TM0(±1, 0), from shorter to longer wavelengths. Bracketed digit mixes indicate coupling orders. Radiation patterns of scattered light coupling to guided modes are given in **g** and characteristic field distributions of each gSLR modes are presented in **h**. **f** Transmittance spectra mapping for the slab thickness grows from 0 to 1200 nm. Theoretical gSLR mode positions are drawn by colored cures. **g** Radiation patterns of scattered light that contribute to gSLR modes in **e**. **h** Characteristic field distributions of gSLR modes in **e**.

**Experimental validation of gSLRs**

Although theoretical studies have anticipated high Q-factors in SLR metasurfaces, their experimental realization is significantly influenced by fabrication quality, array size, and illumination[43,44]. To experimentally characterize the gSLRs, an AuNSA was fabricated on a quartz substrate using a template method, followed by annealing (Fig. 2a). The fabricated AuNSs have an average diameter of ~ 105 nm and an estimated contact angle of ~ 138°[45]. Subsequently, a thin SU8 polymer film was spin-coated and cured with UV light to serve as the core layer (Supplementary Note 2, sample 1). The array size is 200×200 (120×110 μm$^2$). To investigate the evolution of the gSLR mode with varying core layer thicknesses, layer-by-layer spin-coating and curing processes were employed. The thickness of each growth step is approximately 54 nm.

Transmittance spectra were obtained using a custom-built angle-resolved spectroscopy setup (Method and Fig. S4). Spectral mapping under normal incidence ($\Gamma$ point) is shown in Fig. 2b. The experimental findings are in close agreement with the simulation results (Fig. 1f). When the core layer is thin, the spectrum lacks a narrow resonant peak. As the slab thickness increases, the gSLR peaks begin to manifest according to their mode cut-off thicknesses. Concurrently, the resonant peak positions shift to longer wavelengths due to an increase in the effective indices. Utilizing a rectangular lattice configuration, the TE gSLR modes consistently appear at shorter wavelengths than their TM counterparts under $x$-polarized illumination, and vice versa under $y$-polarized illumination (Fig. S5). The Q-factors and peak heights were extracted by fitting the spectra using the Fano equation (Fig. 2c,d, and Fig. S6)[14]. The TE modes exhibit Q-factors ranging from 159 to 490, whereas those of the TM modes range from 389 to 1489 (Fig. 2d). Notably, this Q-factor is the highest reported for a plasmonic nanoparticle array under incoherent illumination in the near-visible range, particularly given its limited array size[16,38-40]. The lower Q-factors of the TE modes compared to that of the TM modes are attributed to the stronger coupling strength, which also results in larger peak heights (Fig. 2d, lower panel). The coupling strength is primarily influenced by the energies of the scattered light coupled to the guided modes. When the AuNSA is positioned at the bottom, the energies coupled to the TE modes are greater than those coupled to the TM modes (Fig. 1g). To explore the dispersion relationship of the gSLR modes, the angle-resolved spectra of a metasurface with a slab thickness of ~1242 nm are analyzed under both s-polarized (Fig. 2e) and p-polarized (Fig. 2f) incidences. The experimental results correspond well with simulations, and lattice mode characterizations are evident[40,41,43,44,46,47]. In contrast to GMRs, there is a significant splitting between the upper and lower bands of the TE0(0, ±1) mode, indicating strong light-AuNS interactions[42,48,49]. Interestingly, two bound states in the continuum (BIC) modes are found in the TE1(±1, 0) and TM0(±1, 0) gSLR modes under *x*-polarized excitation (Fig. 2e and Fig. S7). It should be emphasized that the TE1(±1, 0) and TM0(±1, 0) gSLR modes with BICs have crossed polarization under *s*-polarized excitation (Fig. S7).

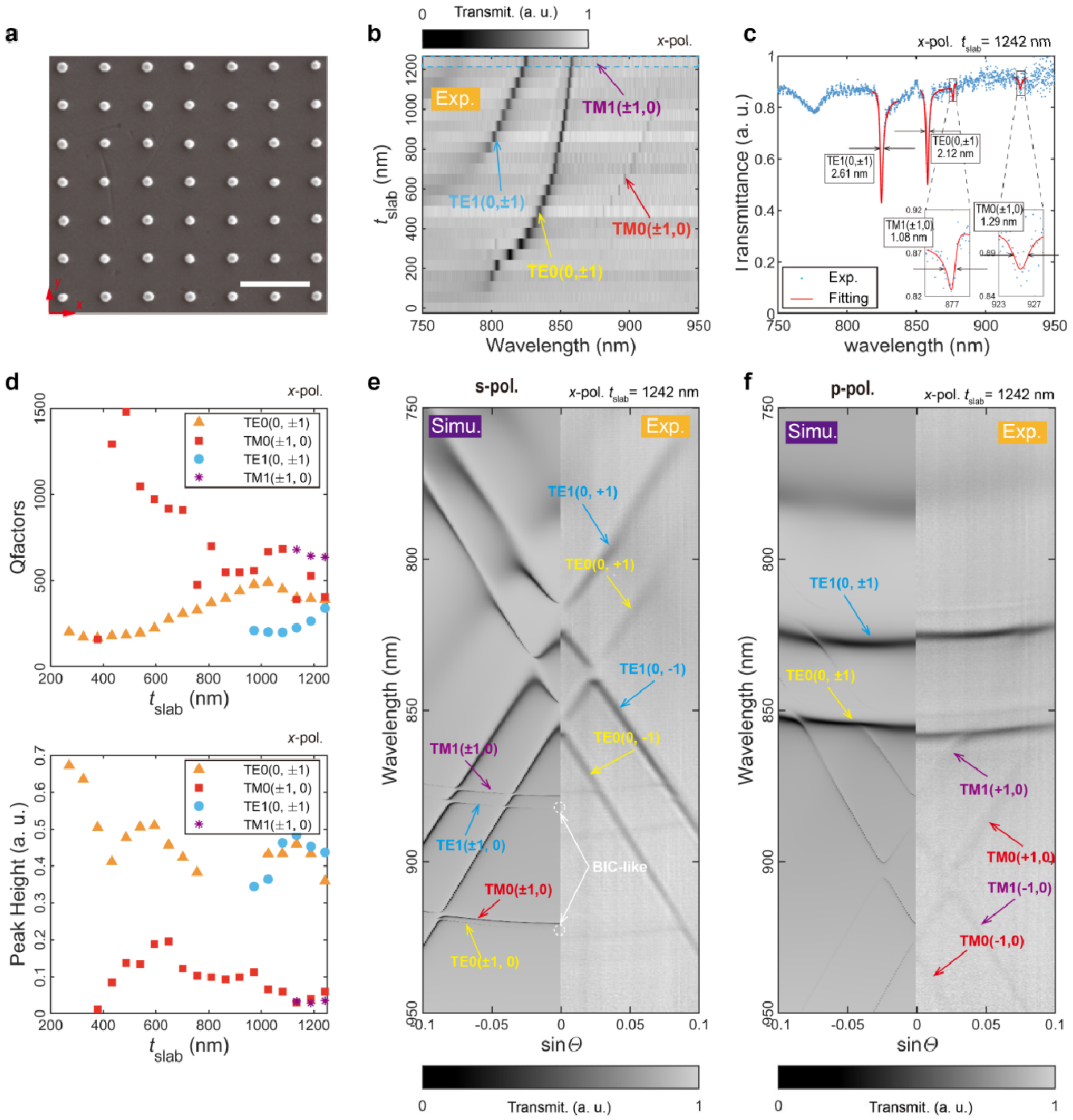

**Figure 2 Experimental realization of gSLRs. a** SEM image of the fabricated AuNS lattice, scale bar, 1 μm. **b** Measured transmittance spectra mapping under *x*-polarized excitation with an increasing slab thickness from 0 to 1242 nm. **c** Experimental transmittance spectrum with slab thickness of 1242 nm. The peak linewidths of the gSLR modes are calculated by fitting the spectrum data with the Fano equation. **d** Q-factors and peak heights of gSLRs as functions of slab thickness. The dispersion features of the gSLR modes are examined by angle-resolved spectroscopy under *s*-(**e**) and *p*-polarized (**f**) illumination. **e** Under *s*-polarized illumination, the TE1(0, ±1) and TE0(0, ±1) modes split, indicating their couplings are in the *y*-direction. Two symmetry-protected BIC modes are found at the Γ point ($\sin\theta = 0$), which degenerate to the TE1(±1, 0) and TE0(±1, 0) modes under oblique excitation. **f** Under *p*-polarized illumination, the TM1(±1, 0) and TM0(±1, 0) modes split, indicating their couplings are in the *y*-direction.

**Mode intensity tuning**

In classic SLRs within an index-continuous medium, the far-field radiation pattern of NPs significantly influences SLR's characteristics, such as coupling directions and mode intensities[12,13,50]. For example, SLRs in a plasmonic nanoparticle array manifest only in the direction perpendicular to the excitation polarization, as resonant plasmonic NPs act as electric dipoles. While in dielectric nanoparticle arrays, SLRs can be excited in both the parallel and perpendicular directions due to the presence of both electric and magnetic dipoles in dielectric nanoparticles[14,51-53]. Analogous to classic SLRs, gSLRs are generated by coupling scattered light carried by the slab waveguides to Bloch modes. Slab waveguides are known to alter the radiation patterns of nanoantennas or emitters within them, with this alteration being position-dependent[54-56]. Consequently, the gSLR mode intensities can be modified by the vertical displacement of the AuNSA within the core layer.

As depicted in Fig. 3a, when the AuNSA is positioned at a vertical displacement $t_{VD}$ from the bottom, the transmittance spectra mapping with $t_{VD}$ varying from 0 to 1200 nm is illustrated in Fig. 3b. From left to right, the four peaks correspond to the TE1(0, ±1), TE0(0, ±1), TM1(±1, 0), and TM0(±1, 0) modes. When the vertical displacement increases, these four modes exhibit variations in peak heights and center wavelengths, implying changes in the coupling strengths. Specifically, the TE1(0, ±1) mode demonstrates strong intensity when the AuNSA is at the bottom, but it becomes negligibly weak for a displacement of approximately 480 nm, followed by an increase and a subsequent decay as the AuNSA approaches the upper boundary. In contrast, the TM1(±1, 0) mode peak has a small height when the AuNSA is at the bottom. then reaches the highest value when the AuNSA is displaced by approximately 480 nm, and goes to a local minumal when the AuNSA is positioned at approximately 810 nm. Minimal changes are observed in the TE0(0, ±1) and TM0(±1, 0) modes before AuNSA contacts the upper boundary (Fig. 3b).

The experimentally measured transmittance spectra with $t_{VD}$ of 0, 514, and 1218 nm in metasurfaces with the $t_{slab}$ of 1218 nm are presented in Fig. 3c (see Supplementary Note 2 sample 2 for fabrication and Fig. S8 for angle-resolved measurements). The experimental results agree well with the simulated ones. The TE1(0, ±1) mode demonstrates a significant peak height the AuNSA is at the bottom, but this peak almost vanishes when $t_{VD}$ is 514 nm and 1218 nm (shaded in green). The TE0(0, ±1) mode exhibits pronounced peaks at $t_{VD}$ values of 0 and 514 nm, with a minor peak when $t_{VD}$ = 1218 nm (shaded in purple). The TM1(±1, 0) mode presents a prominent peak when $t_{VD}$ = 514 nm, with concealed peaks at other positions (shaded in blue). Although the peak heights of the TM0(±1, 0) mode are low across all three positions, a relatively noticeable peak is observed at $t_{VD}$ = 0 (shaded in red).

To investigate the underlying mechanism, we extracted the scattered energies, described by the scattering cross section (SCS), of individual AuNSs positioned at identical vertical locations within a 1218 nm-thick slab (Fig. 3d-g, see Fig. S9 for the SCS spectra over the entire wavelength range). For the TE1 modes, the SCS at 0 vertical displacement is two orders of magnitude larger than those at 514 nm and 1218 nm (Fig. 3d). For the TE0 mode, the SCS at 1218 nm is negligibly small compared to the other two situations. For the TM1 mode, SCS is much greater at 514 nm than at 0 and 1218 nm. Finally, the TM0 mode exhibits the highest SCS value when $t_{VD}$ = 0. The variation trends in SCS values correspond to those in gSLR peak intensities very well, underlining the practicality of precisely tuning gSLR intensities by configuring the vertical displacement of the lattice in the waveguide.

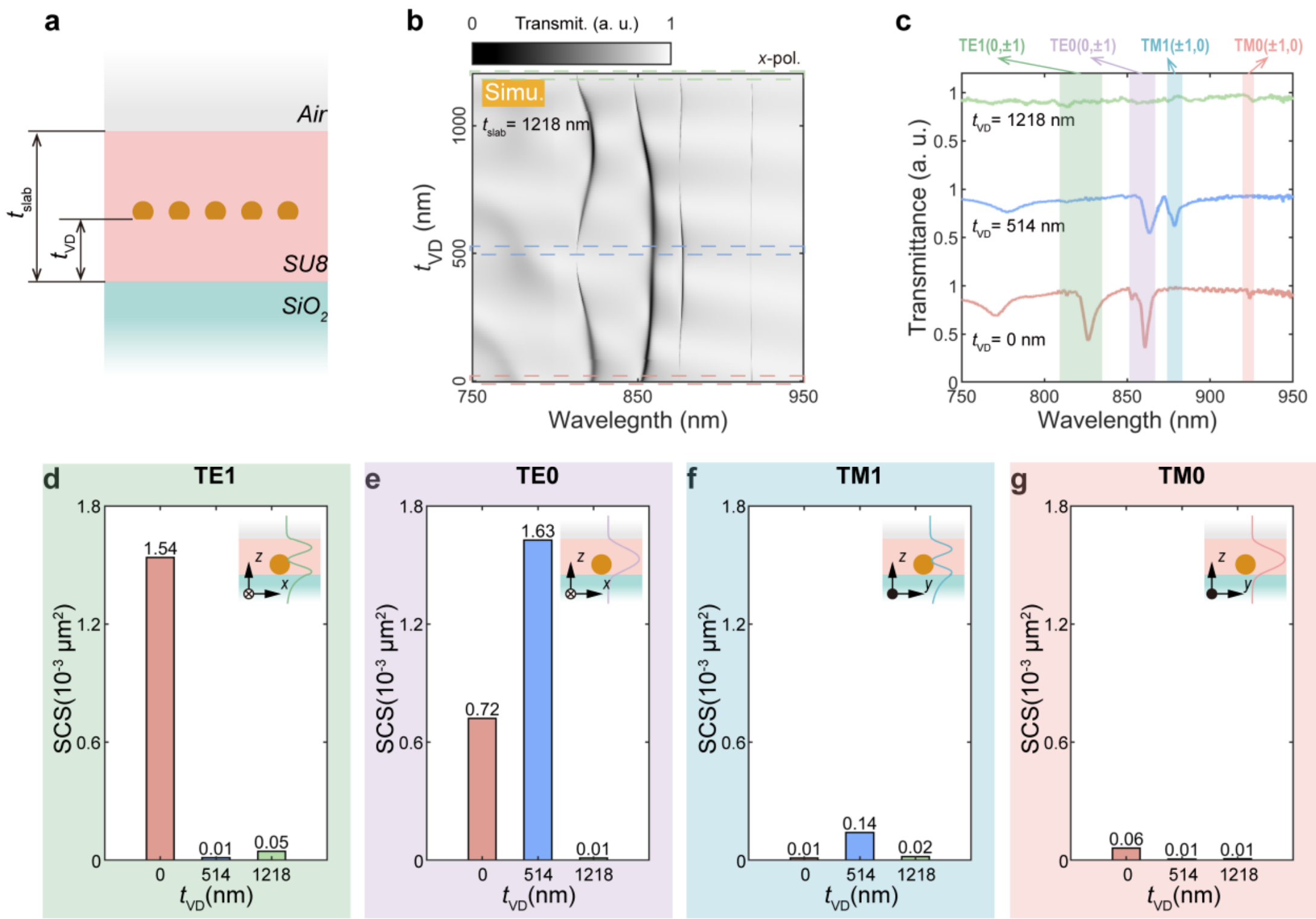

**Figure 3 Tuning gSLRs intensities through radiation pattern modulation by controlling vertical displacement of the lattice in the slab. a** Illustration of a AuNS lattice with a vertical displacement ($t_{VD}$) in the slab. **b** Simulated transmittance spectra mapping of a AuNS lattice in a 1218 nm-thick slab with different vertical displacements under *x*-polarized excitation, with the increase of $t_{VD}$, the intensities and positions of each gSLR modes vary. **c** Measured transmittance spectra of a AuNS lattice in a 1218 nm-thick slab with different vertical displacements of 0 (pink cure at the bottom), 514 nm (blue cure in the middle) and 1218 nm (green cure at the top). Positions of the TE1(0, ±1), TE0(0, ±1), TM1(±1, 0) and TM0(±1, 0) are shadowed by green, purple, blue and pink strips. **d-g** The radiation strengths of corresponding guided modes at different gSLR peak center positions are compared in **d**(TE1), **e**(TE0), **f**(TM1) and **g**(TM0). Inserts are illustrations of guided mode profiles of scattered light. Strong gSLR modes occur at situations when the radiation strengths of AuNSs are large.

**gSLRs in metasurfaces with metal slab waveguides for bio-molecule detection**

As has been introduced above, the development of SLR-based biosensors faces two challenges: the low Q-factors of metasurfaces in index-discontinuous environments and the absence of mathematical models to relate biochemical reactions and optical resonances. Beside generating high-Q resonances, we show gSLRs can address these challenges by deriving a mathematical model that incorporates the biochemical reaction kinetics and the gSLR theory. To improve the sensitivity, we investigate gSLRs in metasurfaces with metal slab waveguides.

The metasurface to be studied is illustrated in Fig. 4b,c (upper panels), which is constructed by placing a gold nanodisk array (AuNDA) on a gold mirror with a dielectric spacer under the gold nanodisks (AuNDs) and covering them with a dielectric capping layer. The fabricated metasurface without the capping layer is shown in Fig. 4a (lower panel, see Supplementary Note 2 sample 3 for details of the fabrication process). The AuNDs have diameters of ~ 180 nm and heights of ~ 45 nm. The lattice constants in the *x*- and *y*-directions are $p_x$ = 600 nm and $p_y$ = 550 nm, respectively. The spacer thickness is ~ 25 nm. When the capping layer thickness is zero, a single AuND sitting on the gold mirror forms a nanoparticle-on-mirror (NPoM ) cavity[57,58]. Scattered light can propagate at the air-gold interface in the form of surface plasmon polaritons (SPPs) (Fig. 4a, upleft), which are also thought to be TM0 modes (Supplementary Note 1). The extracted far-field radiation pattern is in good agreement with the TM0 mode (Fig. 4a, upright). Hence, a TM0(±1, 0) gSLRs mode at ~620 nm is observed in the reflectance spectrum under normal *x*-polarized planewave illumination (Fig. 4b, low). The wide peak at 900 nm corresponds to the localized surface plasmon resonances (LSPRs) of the NPoM cavity. The inset shows the electric-field intensity, and no lattice-mode feature is observed. When a 108 nm SU8 slab is capped on the metasurface, three narrow peaks are found between 500 and 1000 nm (Fig. 4c), which belong to the TE0(0, ±1), TM0(±1, ±1), and TM0(±1, 0) modes according to their field distribution (Fig. 4e, right three). Here, only the field of TM0(+1, +1) is provided (see Fig. S12 for the other degenerated modes). A full view of the gSLR modes evolution against capping thickness is shown in Fig. 4d. Similar to the case of the dielectric slab waveguides, with an increase in $t_{slab}$, a number of gSLR peaks emerge. By plotting the theoretical peak positions of the major modes, a good alignment can be found between the theory and experiments (see Fig. S11 for the simulations).

The fast peak shifting rate with increasing capping thickness suggests that gSLRs in metal slab waveguides are sensitive to changes in dielectric layer thickness on metal surfaces, which can be applied in biosensing. As a proof-of-concept, we validate the feasibility of the BSA detection (Method). BSAs binding on the gold layer surface can be modeled by a uniform dielectric layer. The increase in BSA binding density translates to the growth of the dielectric layer thickness, as illustrated in Fig. 4f (upper panel). This process is finally reflected by the redshifts of the gSLR peak wavelengths. The peak wavelength of the TM0(±1, 0) modes moved by 2.2 nm after reacting with solutions containing 1mM BSA, compared to the phosphate buffer solution (PBS) (Fig. 4f, lower panel). By exploiting the gSLR theory and the biochemical reaction during BSA adsorption, we established a mathematical model that closely combines the biochemical process and optical output (Supplementary Note 3). The relationship between the shift of TM0(±1, 0) mode and BSA concentration can be described by

$$\Delta\lambda = \Delta\lambda_{\max} \frac{c^{n_H}}{K_D^{n_H} + c^{n_H}} , \quad (2)$$

where, $\Delta\lambda$ is the peak shift referred to plain solutions, $\Delta\lambda_{\max}$ is the peak movement at saturated reactions, $c$ is the BSA concentration, $n_H$ is the Hill index and $K_D$ is the equilibrium constant[59,60]. The model fits the experimental results very well, with a curve-fitting degree (R2) of 0.9686 (Fig. 4g). This fitting degree value emphasizes the correspondence between the mathematical model and experiments. The performances of practical biosensors can be promoted by applying advanced manufacture techniques.

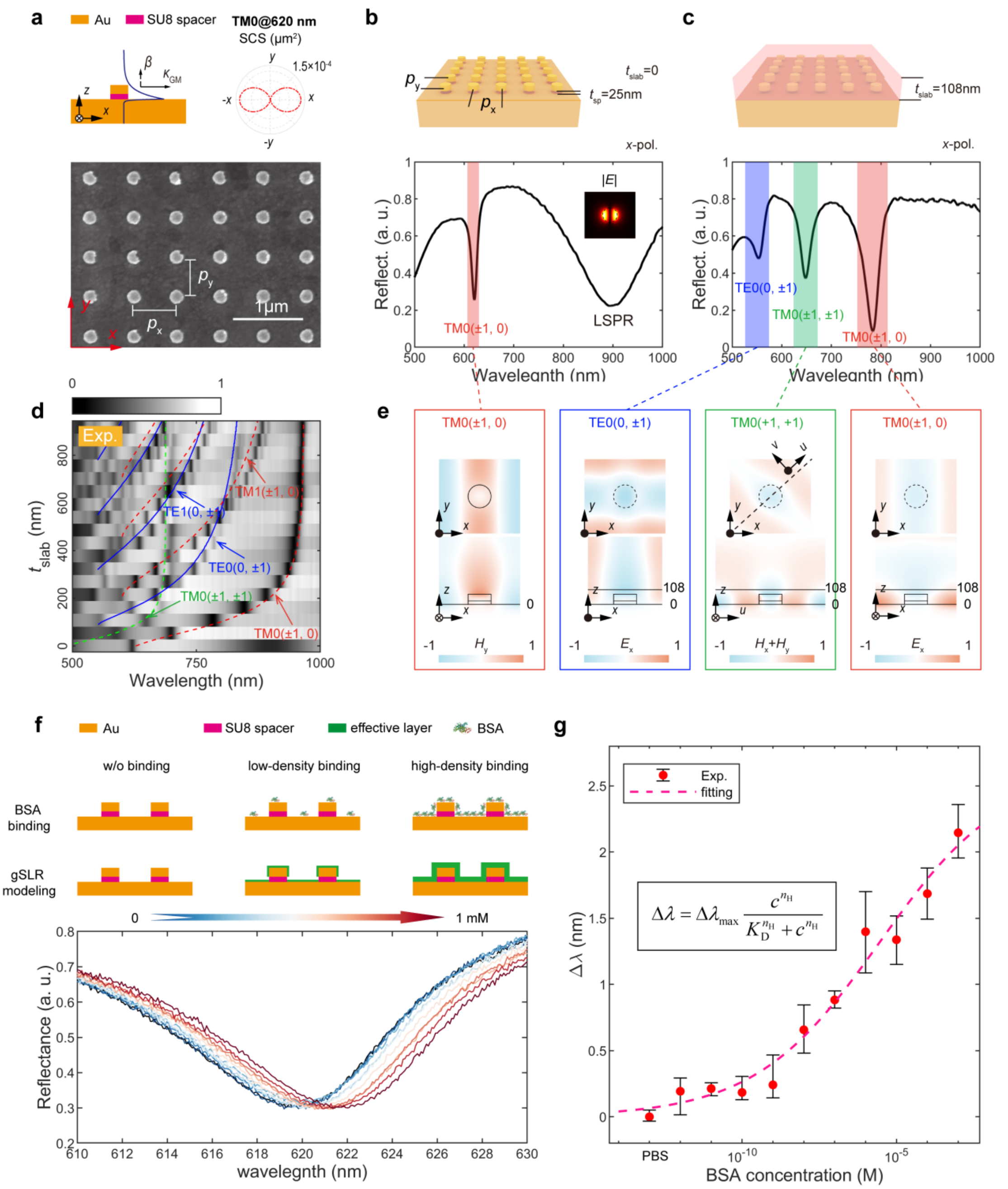

**Figure 4 Characterization of gSLRs in AuNDAs on gold mirrors and proof-of-concept biomolecule detection. a** Schematic showing an AuNDA on a gold substrate and SEM of the fabricated sample. Top left: schematic of an AuND on a gold substrate, scattered light from the AuND couples to the TM0 mode, and propagates at the metal-dielectric interface. Top right: radiation pattern of scattered light couples to the TM0 mode. Bottom: SEM image of the fabricated AuND sample. The diameter of AuND is ~ 170 nm, the height of AuND is ~ 50 nm, the thickness of SU8 spacer is ~ 25 nm, and lattice constants are $p_x$ = 600 nm and $p_y$ = 550 nm, respectively. Experimental reflectance spectra without (**b**) and with (**c**) a 180 nm-thick SU8 slab (**c**) are measured. **b** Without the SU8 slab, the narrow peak is the TM0(±1, 0) gSLR mode and the wide peak corresponds to LSPRs of the NPoM cavities. **c** Three gSLR modes of TE0(0, ±1), TM0(±1, ±1) and TM0(±1, 0) emerge with a 108

nm-thick slab. **d** Experiment reflectance spectra mapping when the slab thickness increases from 0 to 1000 nm. Colored cures are theoretical gSLRs positions. The experiment peak positions match theoretical ones very well. **e** Characteristic field distribution of gSLR modes in **b** and **c**. For clarity, only the TM0( +1, +1) is given, others are given in Supplementary Information. **f** Principle of measuring bio-molecule concentrations by gSLRs. Top: to model the BSA adsorption, the sparsely bound BSA molecules are model by an effective uniform dielectric layer. The BSA density is modeled by the dielectric layer thickness. Bottom: reflectance spectra of the TM0(±1, 0) after reacting with solutions with different BSA concentrations. **g** The relationship of BSA concentration and the TM0(±1, 0) peak center shifts can be fitted by a modified Hill-Langmuir equation.

**Conclusions**

In conclusion, we have demonstrated high-Q multimodal gSLRs in index-discontinuous environments, a Q-factor of 1489 under incoherent illumination, by integrating nanoparticle arrays within slab waveguides. These resonances are generated by coupling scattered light, with radiation patterns modulated by guided modes, to Bloch modes. Leveraging the modulation, we achieved precise tuning of coupling strengths and peak intensities by manipulating the vertical displacements of the nanoparticle arrays within the core layers. In addition, gSLRs exhibit high sensitivities to local dielectric changes, a property we validated for biosensing applications through a proof-of-concept BSA detection. The experimental results were in good agreement with the mathematical model that incorporates both the biochemical reaction kinetics and gSLR theory.

Although the experimental demonstrations and discussions are focused on plasmonic nanostructures, the theoretical framework and techniques presented in this study are applicable to all-dielectric designs, which may lead to higher Q-factor with reduced material losses. Furthermore, the validated sensing mathematical model enables the systematic design and calibration of biosensors for a wide range of analytes, including biomarkers, advancing the clinical application. Beyond biosensing, the demonstrated high coupling efficiency and flexible intensity tunability also highlight the advantages of gSLRs for applications, in which extremely high Q-factors are necessary such as energy harvesting, structure colors, and multiband filters.

## Methods

### Numerical simulations

Finite-difference time-domain (FDTD) simulations were performed to calculate the scattering cross sections of individual nanoparticles and the transmittance spectra of the arrays. For the arrays, simulations were performed for a unit cell using periodic boundary conditions in the in-plane dimensions, and perfectly matched layers in the out-of-plane dimension. A small negative imaginary part ($k = 4\times10^{-4}$) was added to the refractive index of the environment to improve convergence. The free-space and guided radiative patterns of individual nanoparticles were obtained using a finite element method solver (COMSOL Multiphysics) and an open-source algorithm[53]. Momentum-involved eigenmode analysis was carried out in COMSOL with an eigenfrequency solver. Bloch boundary conditions were applied in the in-plane dimensions and perfectly matched layers were employed in the out-of-plane dimension.

### Sample fabrication

Metasurface samples were fabricated using the nanotemple method. Briefly, a silicon nanohole array mold was fabricated by electron beam lithography (EBL) and reactive ion etching (RIE). Subsequently, a poly(diallyldimethylammonium chloride) (PDAC) film was deposited on the mold via spin coating of a 0.5% aqueous PDAC solution. A nanohole array template was fabricated by depositing 2 nm/100 nm Cu/Ag on a PDAC-capped silicon mold and transferring it onto different substrates with the help of water. The AuNDA was prepared by depositing Cr/Au and etching the silver template. Finally, AuNSA was prepared by annealing the AuNDA at 900 °C for 2h. See Supplementary Note 2 for detailed sample fabrication processes.

### Optical measurement

The slab waveguide was fabricated by spin-coating a diluted negative photoresist (SU8-2000.5, 5X in anisole, 5000 rpm to obtain ~ 25-nm-thick films, and 2000 rpm to obtain ~ 54-nm-thick films) on the sample. After coating, the sample was dried on a hotplate at 90 °C for 1 min and cured under 365 nm UV light. The increase in waveguide thickness is realized by a layer-by-layer coating-curing process.

Transmittance spectra were measured using an angle-resolved imaging setup, as illustrated in Fig. S4. Reflectance spectra were measured using another setup, as shown in Fig. S10.

**BSA detection**

BSA was first dissolved in a 1X PBS solution to obtain a 1mM concentration. Solutions of other concentrations were obtained by diluting the prepared 1mM solution. Before detection, the metasurface was treated with UV light for 10 min to remove potential adsorbates. The solutions to be tested were then dropped on the metasurface and left to react for 5 min. Subsequently, the metasurface was rinsed with deionized water (DI) and dried under nitrogen flow. Finally, reflectance spectra were measured using the setup shown in Fig. S10. Experiments were conducted with BSA concentrations from low to high. Repeated tests with the same procedures were conducted to collect the data shown in Fig. 4g.

**Acknowledgements**

S.H., H.W. and Y.P. acknowledge the funding support of State Key Laboratory of Robots and Systems (Grant No. SKLRS202505B). X.D. acknowledges funding from National Natural Science Foundation of China (Grant No. 62275063). Y.Z. acknowledge the funding support of Temple Foundation Endowed Teaching Fellowship in Engineering #2.

**Author contributions**

H. conceived the idea and conducted the experiments. S.H. and K.Y. performed the simulations. S.H., K.Y., H.W. and X.D. analyzed the experimental results. All the authors contributed to the preparation of the manuscript.

**Competing interests**

The authors declare no competing interests.